# Application of Vertex coloring in a particular triangular closed path structure and in Kraft's inequality

Sabyasachi Mukhopadhyay[1]
B.Tech 3rd Year, ECE
College of Engineering & Management, Kolaghat
KTPP Township
Dist: East Midnapur, India
sabyasachi.unique@gmail.com

Dr. Paritosh Bhattacharya[2]
College of Engineering & Management, Kolaghat
KTPP Township
Dist East Midnapore, India
p_bhattacharya2001@yahoo.co.in

Prof.(Dr.) B.B.Ghosh[3]
Professor, IEM Salt Lake
Ex-Professor IIT Kharagpur, India
bankimghosh@hotmail.com

***Abstract*— A good deal of research has been done and published on coloring of the vertices of graphs for several years while studying of the excellent work of those maestros, we get inspire to work on the vertex coloring of graphs in case of a particular triangular closed path structure what we achieve from the front view of a pyramidal structure. From here we achieve a repetitive nature of vertex coloring in case of odd and even number of horizontal lines within this triangular structure. In order to apply this repetitive nature of vertex coloring in case of a binary tree, we get a success in Kraft's Inequality. Actually our work mainly deals with a particular triangular closed path vertex coloring and repetition of the vertex coloring nature in case of the Kraft's inequality in the field of Information Theory and Coding.**

## I. INTRODUCTION

The first results about graph coloring deals almost exclusively with planar graphs in the form of the coloring of maps. While trying to color a map of the counties of England, Francis Guthrie postulated the four color conjecture, noting that four colors were sufficient to color the map so that no regions sharing a common border received the same color. Guthrie's brother passed on the question to his mathematics teacher Augustus de Morgan at University College, who mentioned it in a letter to William Hamilton. Then Arthur Clayey raised the problem at a meeting of the London Mathematical Society. The same year, Alfred Kempe published a paper that claimed to establish the result, and for a decade the four color problem was considered solved. For his accomplishment Kempe was elected a Fellow of the Royal Society and later President of the London Mathematical Society.Heawood pointed out that Kempe's argument was wrong.Heawood himself modified that thought. This background history & thoughts of those maestros has inspired us to predict our examined vertex coloring property in case of a particular triangular closed path structure and in Kraft's Inequality satisfying prefix coding .

## II. MATHEMATICAL MODELING

While dealing with the pyramidal structure, we get a front view of a triangle shaped Structure. This shown in the following figure-1.(a)

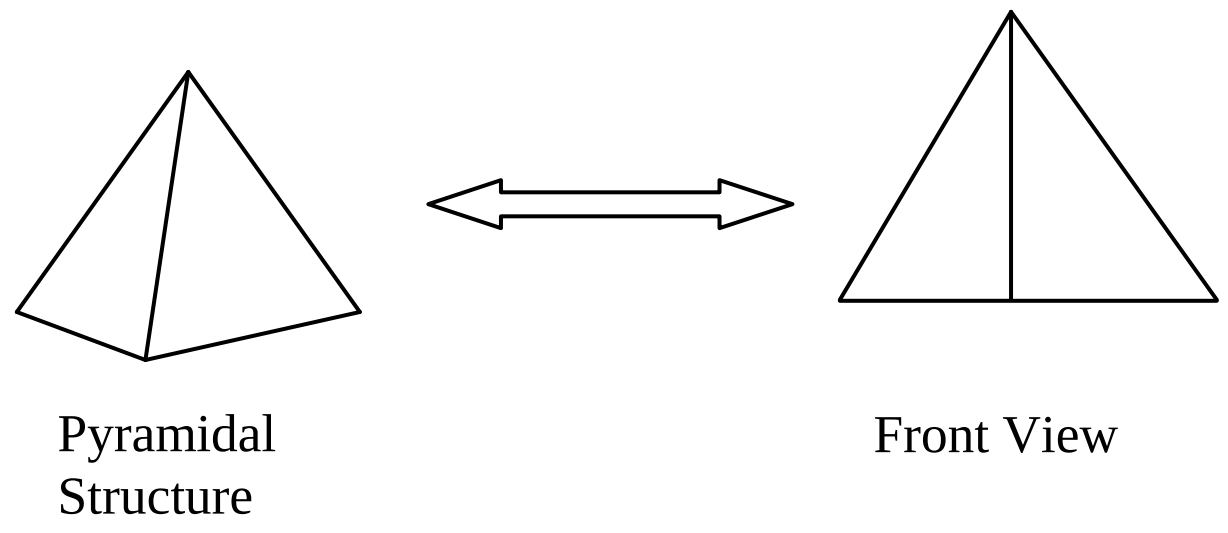

**Fig.1.(a)**

Thereafter experimenting with the above figure, we get a particular pyramidal structure with its front view shown in the figure-1.(b)

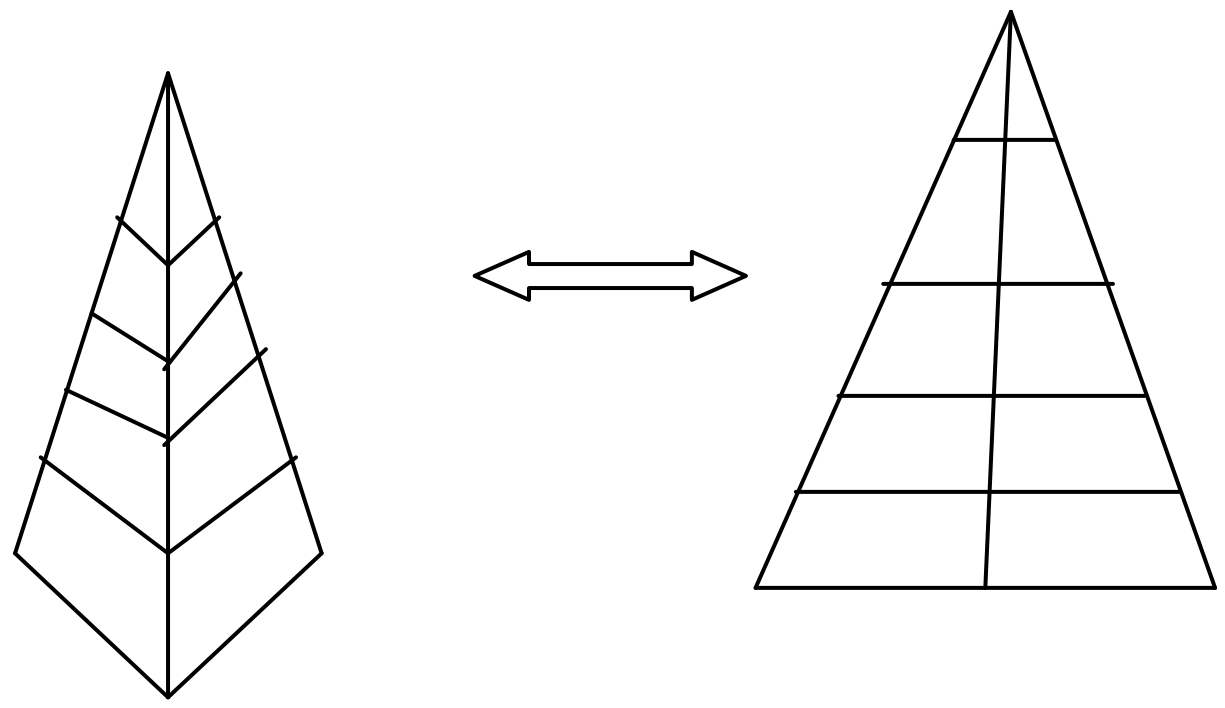

**Fig.1.(b)**

Proceeding in this way we consider a case where we have to place 91points in such a way that 10 points exist in each of the 10 inclined lines and in each of the 9 horizontal lines too. How will we arrange this?

The solution is shown in the following figure-1.(c)

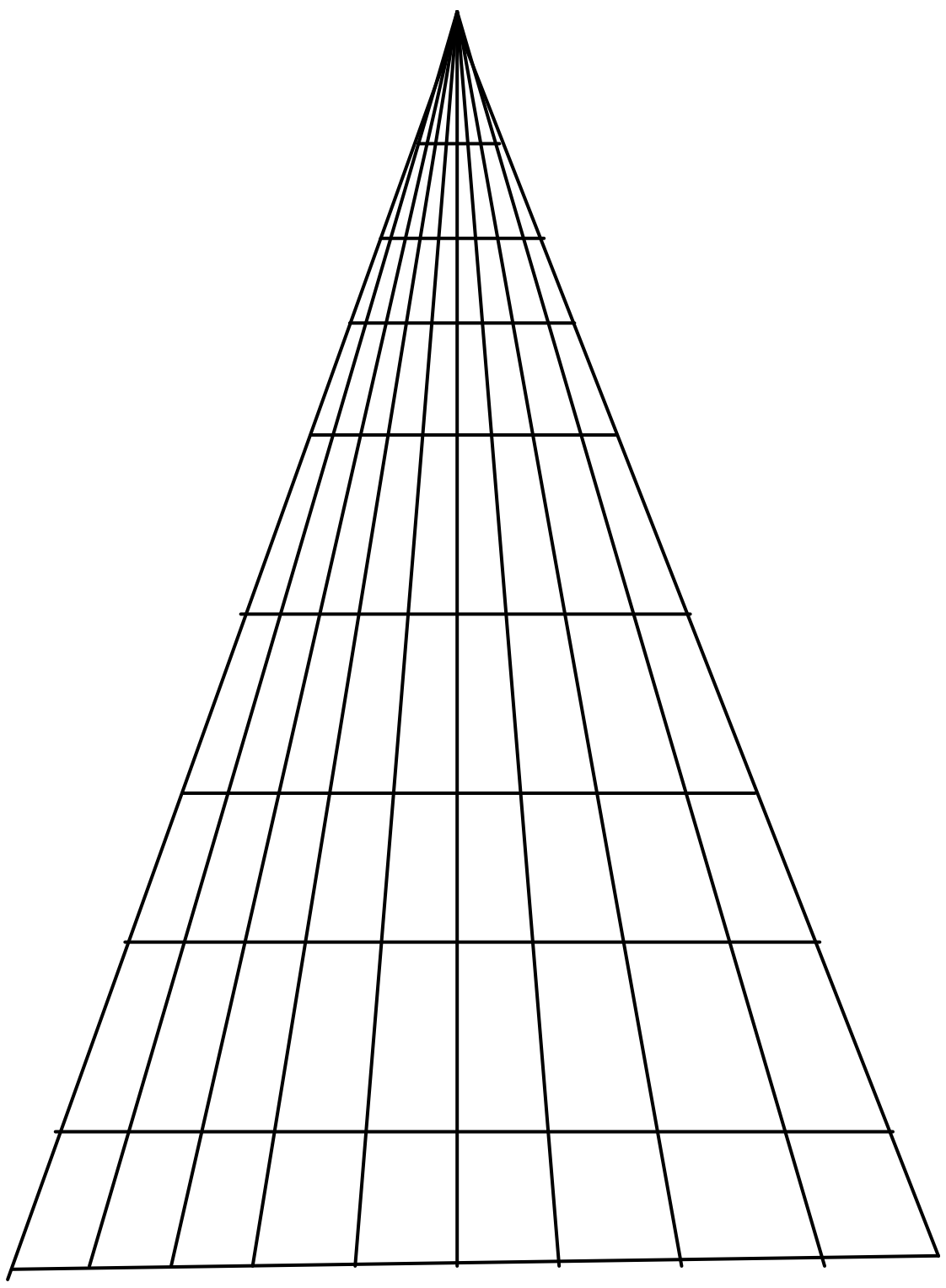

**Fig.1.(c)**

Now let's have a look at the small demo of the same type of problem where arranging 7 points where 3 points exist in each of the 3 inclined lines and each of the 2 horizontal lines too. This is shown in the following figure-1.(d)

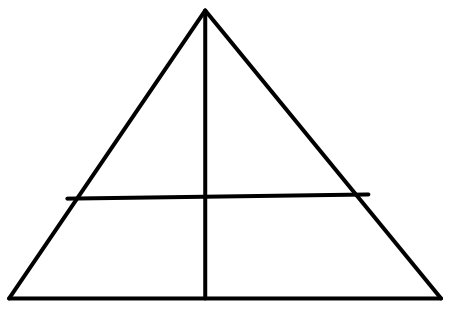

**Fig.1. (d)**

From these above cases we have come to an important general assumption by mathematical induction that if we arrange n points in each of n number of inclined lines and each of (n-1) number of horizontal lines too, then we can arrange total number of [n(n-1)+1] points in this way.

Now if we want to apply a coloring property in these cases then we can see the following results.

We know that in case of vertex coloring to paint the vertices of a graph G, no two adjacent vertices have the same color as well as in case of edge coloring to paint the edges of a graph G no two adjacent edges have the same color.

Now here we have applied vertex coloring in case of the following structure shown in Figure .2.(a).

Here we arrange 13 points where 4 points in each of the 4 inclined lines and in each of the 3 horizontal lines.

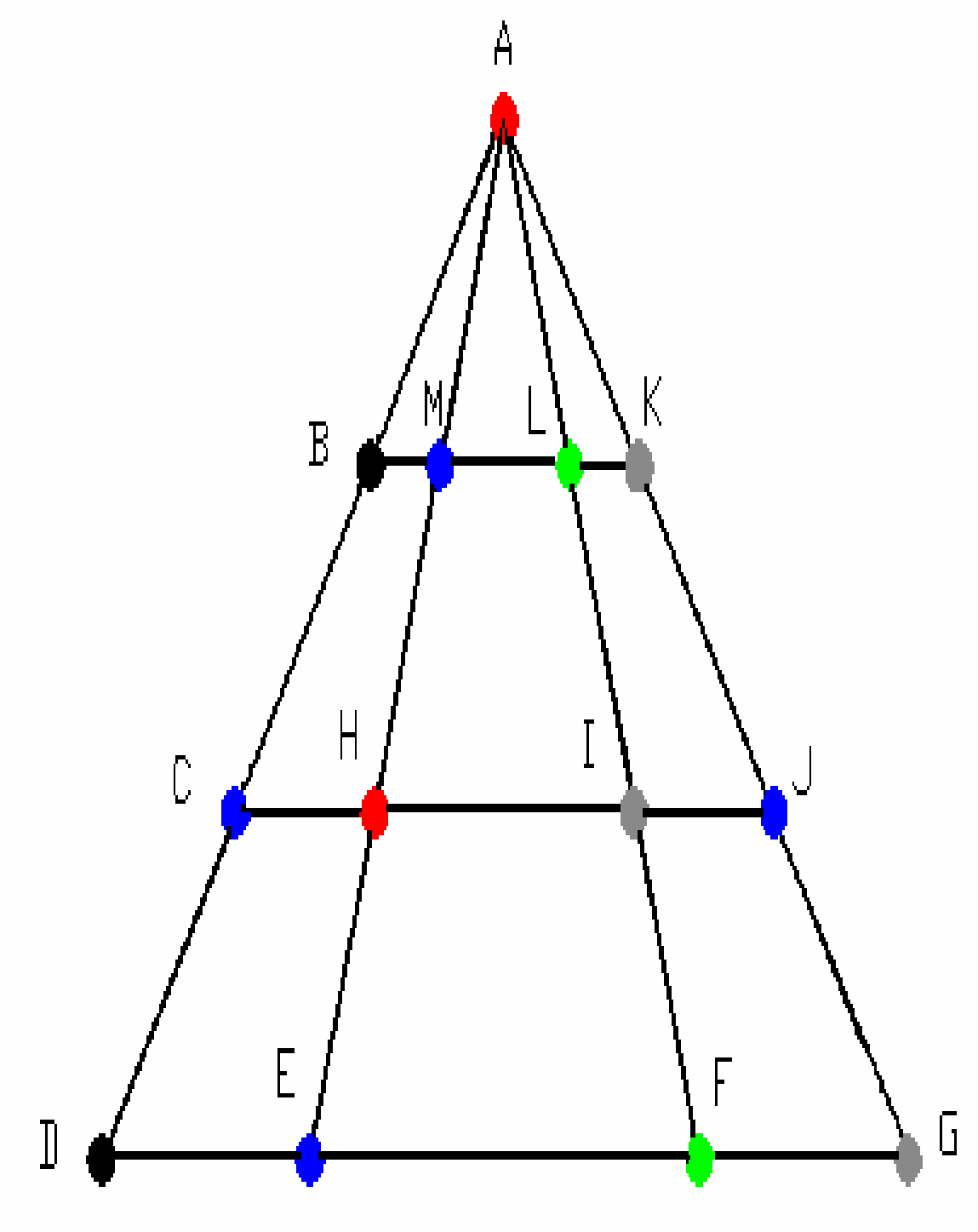

**Fig.2. (a)**

If we color A by red, then due to the property of coloring of graph (B, M, L, K) we consider any other color apart from red, is black, blue, green and grey respectively. Keeping the vertex coloring property in mind, we also represent (C, H, I, J) points in color with blue, red, grey and blue respectively. The last row (D, E, F and G) points are colored in black, blue, green and grey. Other combination of coloring following partition theory can also be applicable but here we can notice the vertex color combinations in 1st and 3rd horizontal lines are same.

Let see another example for representing 31points where 6points in each of the 6 inclined lines and in each of the 5 horizontal lines. This is shown in the following figure-2.(b)

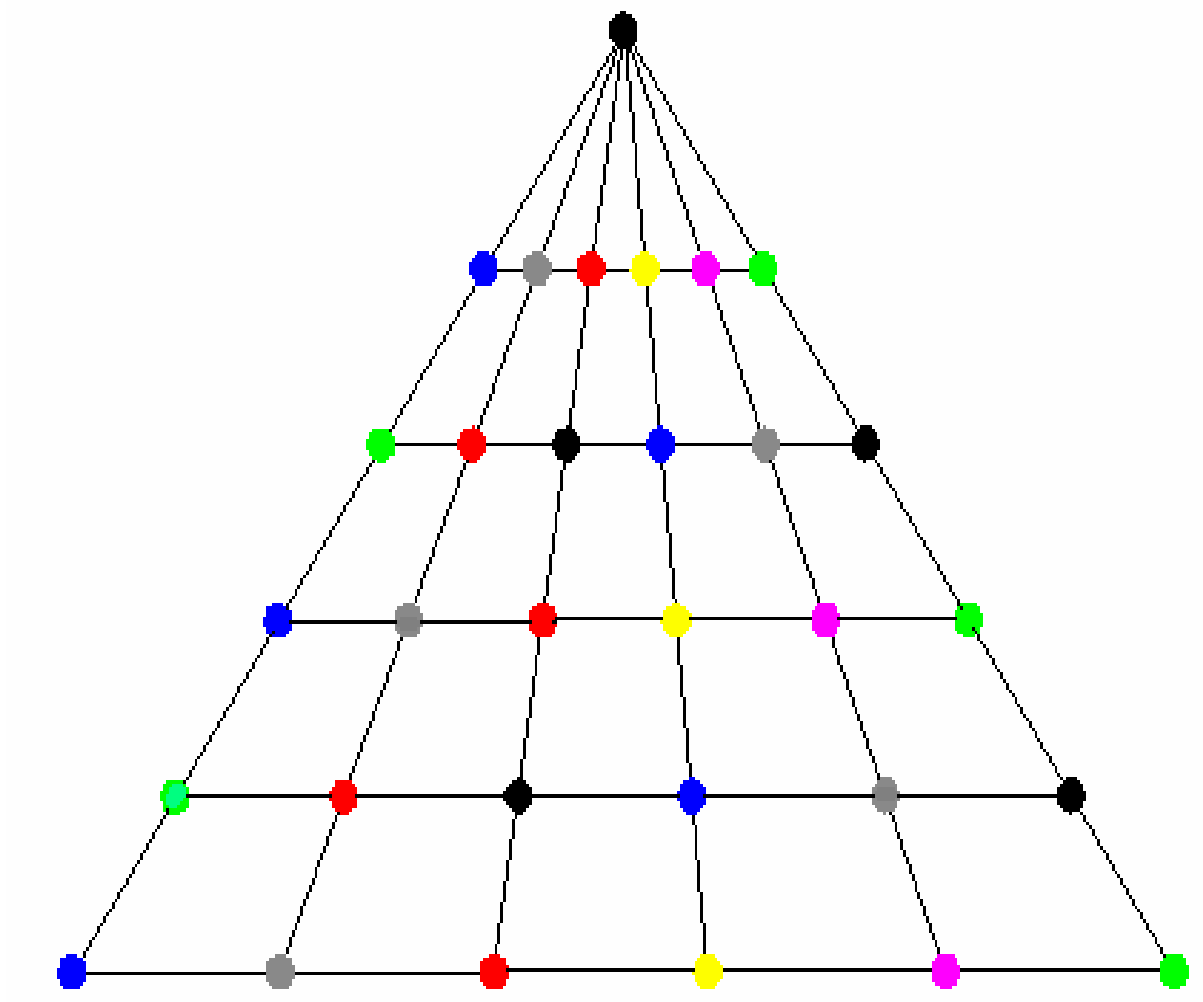

**Fig.2. (b)**

Here we notice that within this particular triangular structure , the vertex color combinations of 1st,3rd and 5th horizontal lines are same and also the vertex color combinations of 2nd and 4th horizontal lines are same without breaking vertex coloring principle.So,our approach is successful in case of triangular closed path structure by applying vertex coloring.

Therefore by taking mathematical induction for that type of the particular triangular structure due to the vertex coloring that the vertex color combinations of (2n-1)th horizontal lines and also the vertex color combination in 2n th horizontal lines are same without breaking vertex coloring principle.

## III. RESULTS AND DISCUSSIONS

We have seen the vertex coloring repetition incident in the above discussion in case of odd and even number of inclined and horizontal lines within a particular triangular closed path structure. The success in the above case inspires us to work on the further extent. So we will apply this coloring repetition property in case of the binary trees to explain the Kraft's inequality.

Kraft's inequality states that a necessary & sufficient condition for the existence of a binary code with code words having lengths $n_1 \leq n_2 \ldots\ldots\ldots\ n_L$ that satisfy the prefix condition is

$$\sum_{k=1}^{L} 2^{-n_k} \leq 1$$

Where a prefix code is one in which no codeword forms the prefix of any other codeword.

Applying coloring property using a binary tree for the construction of a prefix code
is shown in below figure- 3

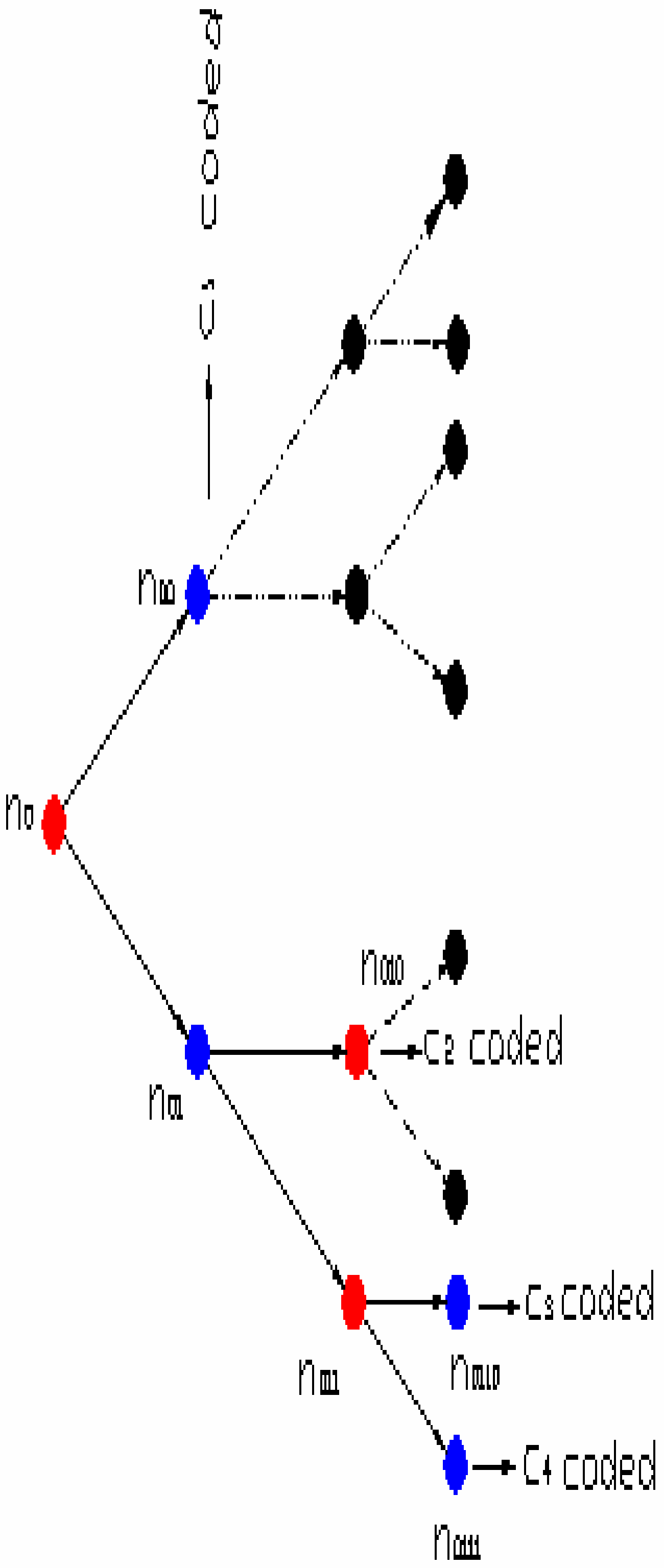

**Fig.3**

From the definition of proper vertex coloring we know that it will be efficient only when we use minimum number of color in case of vertex coloring successfully. So in this case for the coloring of the nodes of a binary tree we use only two colors such as red and blue respectively.

Now we start from mother node colored by red (by assuming) and approach towards the terminal nodes.

There are two branches in binary tree. Let both of the upper and lower branch are colored as blue. At 1st we follow the upper branch from the mother node $n_0$.As a result we obtain the 1st codeword $C_1$ and terminating that blue colored node $n_{00}$ according to the definition of prefix code. Now in case of lower branch from the mother node $n_0$, we reach the blue colored node $n_{01}$.We approach the upper branch 1st and reach its red colored node $n_{010}$ (here we assume the color of both child node of blue colored node as red in order to show the repetition coloring in next steps).That red colored node $n_{010}$ is coded as $C_2$.Follwing the lower branch from blue colored node $n_{01}$, we ultimately reach the terminal nodes $n_{0110}$ and $n_{0111}$, which are both colored as blue. Here both of them are coded as the code words $C_3$ and $C_4$ respectively. Hence we see the coded colored for the starting code $C_1$ and the terminating codes $C_3$ and c4 are same i.e., blue colored. Hence again repetition of coloring is successfully taken place in this case. Now we all know

$$\sum_{k=1}^{L} 2^{-n_k} = 2^{-1} + 2^{-2} + 2^{-3} + 2^{-3}$$

$$= 0.5 + 0.25 + 0.125 + 0.125 = 1$$

Hence Kraft's Inequality is satisfied. So over all we see our main objective is successfully applied in case of the prefix coding for Kraft's inequality.

## IV. CONCLUSIONS

Generally we see the approach of vertex coloring is limited within the case of trees in graphs but here we use this vertex coloring in case of a particular triangular closed path structure. As a result of it, we observe the vertex color repetition in case of the odd and even horizontal lines within a particular triangular closed path structure. We expect it will create a new era in the study of graph theory and the researchers will be influenced to work on it in future in the field of Image Processing and Digital Signal Processing. Our another approach of vertex coloring on binary trees for the explanation of Kraft's inequality by prefix coding is also successful. We hope this will also create a new era to think the researchers on the development of this particular pattern in the case of Information Theory and Coding. Overall our approach belongs to the field of Digital Communication, Computer Science and Information Technology oriented.

## ACKNOWLEDGEMENT

We wish to acknowledge Mr. Saradindu Chakraborty of College of Engineering & Management, Kolaghat for his help in case of drawing the vertex colored figures.